\documentstyle[epsfig, twocolumn]{iso98}

\newcommand{\thepapername}{}

\begin{document}

\setlength{\parindent}{0pt}
\setlength{\parskip}{ 10pt plus 1pt minus 1pt}
\setlength{\hoffset}{-1.5truecm}
\setlength{\textwidth}{ 17.1truecm }
\setlength{\columnsep}{1truecm }
\setlength{\columnseprule}{0pt}
\setlength{\headheight}{12pt}
\setlength{\headsep}{20pt}
\pagestyle{veniceheadings}

\title{\bf The European Large Area ISO Survey\\
ISOPHOT results using the MPIA-pipeline}

\author{{\bf C. Surace$^{1,2}$,P. H\'eraudeau$^1$, D. Lemke$^1$, S. Oliver$^2$,
 M. Rowan-Robinson$^2$ and the ELAIS consortium$^3$} \vspace{2mm} \\
$^1$ Max Planck Institut f\"ur Asrtonomie, Heidelberg, Germany \\
$^2$ Imperial College of Medecine Science and Technology, London, UK\\
$^3$ European consortium involving 19 countries}

\maketitle

\begin{abstract}
The European Large Area ISO Survey (ELAIS) will provide Infrared observations of 
4 regions in the sky with ISO. Around 2000 Infrared sources have been detected at 
7 and 15 $\mu$m (with ISOCAM), 90 and 175 $\mu$m(with ISOPHOT)) 
over 13 square degrees of the sky. 
We present the source extraction pipeline of the  90~$\mu$m ISOPHOT observations,
describe and discuss the results obtained and derive the limits of the ELAIS observation 
strategy. 
        \vspace {5pt}\\


  Key~words: ISO; ISOPHOT; Survey; Methods: data-analysis.

\end{abstract}

\section{INTRODUCTION}
ELAIS (P.I.: M. Rowan-Robinson) is a 
consortium involving 19 European institutes. The project consists on an ISO
survey of 13 square degrees of the sky distributed mainly in 4 areas 
(3 in the northern hemisphere, 1 in the southern one). 
These areas have been chosen to be at high ecliptic latitude, to have low cirrus
bacgkround, and to avoid the presence of strong 12 $\mu$m sources closeby 
(\cite{so98}). We used the mapping mode of ISOPHOT and the C\_100 detector.
 Each area was initially planned
to be covered by a $20 \times 10$ raster map without overlapping. 
Details on the observing modes and the characteristics of 
the survey can be found in \cite*{so97}. \\
Becasue ELAIS is the largest open time project (around 360 hours) being undertaken by 
ISO, the regions observed will be one of the most well studied at several wavelengths
and will be some of the most adequate regions to explore the extragalactic universe.\\

To treat the large amount of data several data extraction pipelines have been 
set-up at Imperial College (London), IAS and CEA (Paris) and MPIA (Heidelberg).
These pipelines provide the basic data for ground based telescope follow-ups at
different wavelengths (\cite*{ph99}, \cite*{pc98} (see \cite{mrr99} for a review))
and essential information to constrain the spectral energy distribution in the Mid
and Far Infrared Band.  \\
In the following sections I will describe the ISOPHOT MPIA data extraction 
pipeline and discuss quantitatively the results obtained.   

More information about the status of the survey and ELAIS ground-based follow-ups 
is available at the WEB address: {\it http://athena.ph.ic.ac.uk/}.

\section{Data analysis}
\label{sec:data_anal}
\subsection{Data raw analysis}

The raw data have been analyzed using the iso{\b P}hot {\b I}nteractive {\b
A}nalysis tool (\cite{cg96}) (PIA\footnote{PIA is a joint development by the ESA
Astrophysics Division and the ISOPHOT consortium} Version~7.2.2).  We use PIA to
reduce the raw data from the ``Edited Raw Data (ERD)'' level to the ``Auto Analysis
Product (AAP)'' level, keeping the integrated values during the ISO slewing mode.
For deglitching we used the very low sigma value of 1.5 for the sigma-deglitching in
order to get rid of the fluctuations of the detector after being hit by a
glitch.  We use the FCS measurements to calibrate the data.  The calibration was
checked using dedicated measurements (see section \ref{sec:calib}) We used
private routines (\cite{cri97}) to reduce the pixel-to-pixel response fluctuations
and to derive the final surface brightness values and fluxes of the objects.

\subsection{Calibration}
\label{sec:calib}

\begin{figure}[!h]
  \begin{center}
    \leavevmode
    \centerline{\epsfig{file=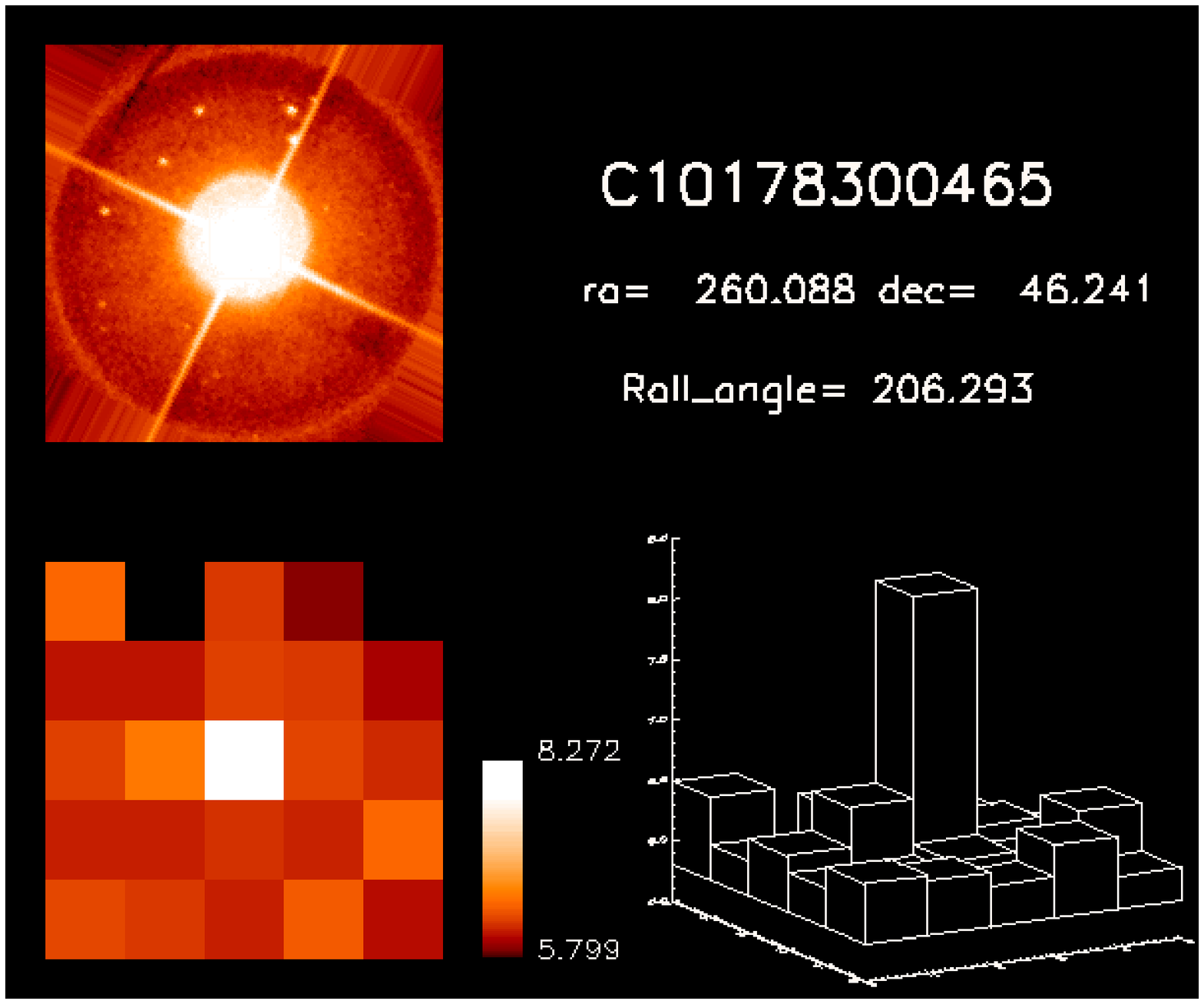,width=6cm}}
    \centerline{\epsfig{file=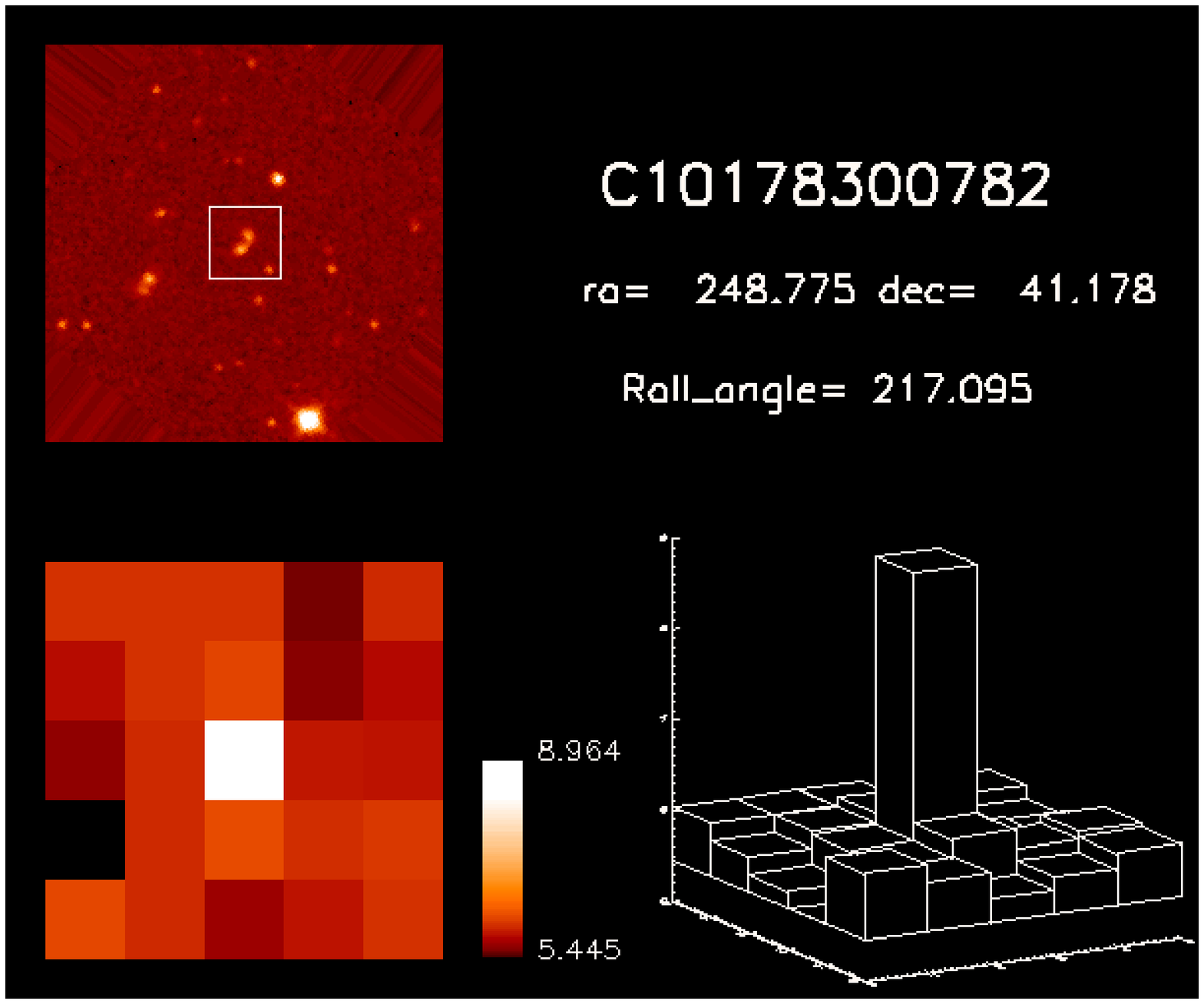,width=6cm}}
  \end{center}
  \caption{\em Two calibration dedicated measurements. 
  Each frame is composed by the DSS image (upper left, rotated by the value of the
   roll angle) the ISOPHOT C100 image (lower left, intensity range is display besides
   in MJy/sr) the 3D view of the total map (lower right corner).
   The upper frame is the standard star, the lower one is one ISOPHOT source.}
  \label{fig:calib}
\end{figure}

The calibration step of the ISOPHOT data is still on discussion despite of the
constant progress in building good calibration files.  We decide to check the
quality of the FCS calibration of the survey observing calibration stars and
sources detected in the ELAIS survey.  The strategy is to observe alternatively
a standard star and an ELAIS source, in each field and to compare the fluxes
obtained in these measurements with theoretical values (\cite{bs98}) and the
flux values derived previously from the survey.  The measurements have been
performed using the ELAIS observing mode for better consistency.  the fig
\ref{fig:calib} shows exemples of observations performed for that purpose.

We found that the fluxes derived in our pipeline where in relatively good
agreement for the point-like sources:

\begin{itemize}
 \item{The fluxes derived using the FCS calibratiton and the theoretical values
agree within 15\% of the total fluxes.}
\item{The fluxes derived during calibration measurements agree within 5\% with
those derived from the survey}
\end{itemize}

\subsection{Target detection} 

\begin{figure}[!h]
  \begin{center}
    \leavevmode
    \centerline{\epsfig{file=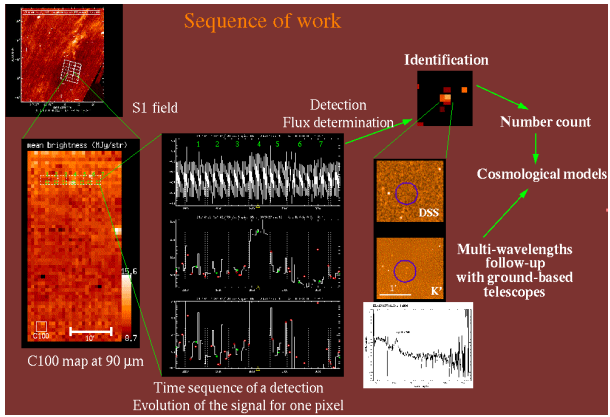,width=8cm}}
  \end{center}
  \caption{\em Summary of the data extraction procedure. From the ISO 
  observation, the detection is performed on the time sequence.  This gives the
  basic information for the identification of the sources and the follow-ups.  }
  \label{fig:sow}
\end{figure}

The non overlapping mapping mode of the ELAIS survey makes impossible the use of
the built-up map for detection purpose.  The noise resulting in building the
map will not allow an efficient detection.  The detection of the ISOPHOT
sources is based directly on the time sequence analysis of the AAP files.  The
analysis process is summarized in fig \ref{fig:sow}.  The basic idea of the
pipeline is to detect the maximum of objects by a 1.5 sigma threshold search
over the time sequence of the ratio of one pixel over the other 8 pixels of the
C100 detector.  The spurious detections are rejected by a second threshold
search on the redundancy.  The pipeline can be summarized by the different steps:  
\begin{itemize} 
\item {ratio by raster point of each pixel over the 8 others
of the C100 pixels (8 values for 1 pixel)} 
\item {1.5 sigma threshold detection
over the time sequence of a pixel (median over 5 raster points)} 
\item {1.5
sigma threshold detection over the time sequence of the ratio for each pixel
(median over 5 raster points and fit of the background over 5 raster points)}
\item selection of the pixels the sum of the detections of which is greater than
4.  
\item{computation of the flux of the detection (flux of the detection pixel
- median(flux of 26 surrounding pixels)) } 
\item{computation of the coordinates of the detections} 
\end{itemize}

The detections are visually inspected by 5 differents people
 and classified in 5 groups define as follows :

- class 1: Clear detection of a bright source \\    
- class 2: Clear detection of a source     \\
- class 3: probable detection of a source    \\
- class 4: affected by glitches, not a source  \\   
- class 5: spurious detection     \\

A final classification is given to the detections.  In the following we will
only consider the detections belonging to the classes 1, 2 and 3.  These
detections represent approximatively 10\% of the total number of detections.

The closeby ($separation < 50"$) ``real'' detections are groupped and the
corresponding fluxes are added.  The data are saved in an IDL structure format
and include in an IDL database.  An ASCII table is also provided, with:

- {\it num:} numero of the detection in the raster \\
- {\it $\alpha$:} right ascension (J2000)\\
- {\it $\delta$:} declination (J2000)\\
- {\it flux:} flux in mJy\\
- {\it sigma:} uncertainty in mJy\\
- {\it det:} detection level ($1 < det < 10$)\\
- {\it time:} central time of the detection in seconds\\
- {\it pix:} pixel number ($0 < pix < 8$) \\
- {\it rp:} raster point ID of the detection\\
- {\it classif:} classifcation given to the detection.\\

see table \ref{tab:data} for an exemple.

\begin{table*}[htb]
  \caption{\em exemple of table generated by the MPIA pipeline.}
  \label{tab:table}
  \begin{center}
    \leavevmode
    \footnotesize
    \begin{tabular}[h]{cccccccccl}
      \hline \\[-5pt]
       num &    $\alpha$  (2000)& $\delta$  (2000) &   flux  &    sigma & det 
       & time  & pix &rp &classif.\\[+5pt]
      \hline \\[-5pt]
 13 &  0 31 44.77 &-42 32 50.07 & 117.2823  & 31.6307 & 4 & 601.930 & 2 & 19 &2/2/2/2/2/\\
 26 &  0 31 34.37 &-42 44 20.68 & 214.7283  & 41.2120 & 9 &1066.973 & 2  &34 &1/2/1/1/1/---1\\
 27 &  0 31 35.27 &-42 45  6.11 & 132.6574  & 41.3140 & 9 &1066.973 & 5 & 34 &2/2/2/2/1/---1\\
 56  & 0 30 52.44 &-42 43 10.48 &   8.4550  & 22.1327 & 5 &1966.065 & 7 & 63 &3/3/3/3/3/\\
      \hline \\
      \end{tabular}
  \end{center}
  \label{tab:data}
\end{table*}

The optical identification is performed on the DSS images, and R-band survey when
 available, using a maximum likelyhood method.

\section{Results and limits}
\subsection{detection limit}

\begin{figure}[!h]
  \begin{center}
    \leavevmode
    \centerline{\epsfig{file=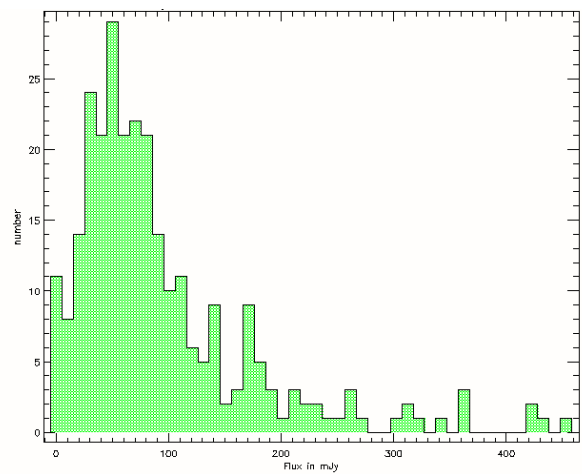,width=8cm}}
  \end{center}
  \caption{\em Distribution of the fluxes of the detections, derived using the 
  pipeline.}
  \label{fig:histo_flux}
\end{figure}

The figure~\ref{fig:histo_flux} shows the distribution of the fluxes derived
from our pipeline.  Even if some sources have been detected down to 10 mJy.  The
uncertainties of the fluxes is found between 20 and 60 mJy, depending on the
observation itself.  The detection of some sources with a signal to noise value
less than one is due to the fact that the detection and the computation of the
flux and uncertainties are based on two different processing modes.  While the
detection is performed on each individual pixel the computation of the flux is
derived on all the pixels of the detector.  The sigma values on the fluxes
include also the pixel-to-pixel fluctutaions.

Nevertheless, some rasters have been lost due to the fact that they have been
observed during the end of the orbit.  Indeed the resulting noise is too high to
be able to detect any source contribution.

The drop off of the histogramme indicates a detection limit at 60 mJy.  While
the median is 80 mJy and the completness of the survey is reached for a limit of
100~mJy (\cite{mrr99})

One has to be aware that due to this detection limit, some objects which are
brighter than 100 mJy could be missed by the survey.  Indeed if, due to the
position of an object in the field, its flux could be distributed among at least
two neighbouring pixels, and then not be detected by the pipeline.  We estimate
to miss 20\% of these objects in the survey.  This situation improves when using
the overlapping observational (overlap 3/2 pixels) mode.

\subsection{Spatial distribution}
\label{sec:spa_dis}
Up to now 45 fields have been analyzed leading to a total of 8164 detections 
that have been eye-balled. The observers have agreed on a number of 274 
``reliable detections'' (class 1 or 2)) and 642 ``probable sources'' (class 3).
One has to remind however that a bright object could be detected several times. 
This leads to a density of 30 detections by square degrees.
The large number of ``class 3'' object is due to the behaviour of the C\_100 
detector, when hit by glitches. A forthcoming new deglitching method will
improve the reliability of the sources.

\begin{figure}[!h]
  \begin{center}
    \leavevmode
    \centerline{\epsfig{file=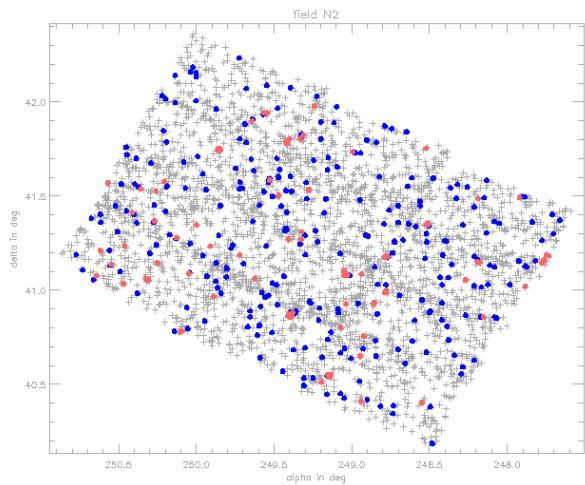,width=8cm}}
  \end{center}
  \caption{\em Spatial distribution of the ISOPHOT detections.
 The light crosses are the detections of possible sources.  The dark filled
circles are the ``class 3'' detections, the light filled circles are the ``class 2''
detections}
  \label{fig:detec}
\end{figure}

The fig~\ref{fig:detec} shows the spatial distribution of the detections on the
N2 ELAIS field.  The light crosses show the detections by the pipeline.  The
dark filled circles are the so-called ``class 3'' detections while the light
filled circles are the ``class 2'' detections.

The overdensity of the detections in some area is an artificial effect due to the 
overlapping regions between two rasters.
There is no clear evidence of some 2D clustering properties of the "real sources".
They look smoothly distributed in the field. 
However the ``class 3'' detections are often linked to the ``class 2'' detections. 

This could be explained by the fact that:\\
- the source could have a morphology leading to a bright center and fainter outer parts,\\
- the source is not centered in the pixel so part of the flux is in the neighbouring pixels \\
- the beam profile of bright sources , could affect as well the neighbouring pixels\\

A more elaborate analysis taking into
account the signal to noise ratio of each raster as well as the overlapping regions
is ongoing.

\section{Conclusion}
\label{sec:conc}
In this contribution we have presented the MPIA ISOPHOT ELAIS pipeline. The results from this pipeline 
will be cross checked with those obtained with other pipeline, in order to get a complete source list
to deliver to the scientific community. We have cross-checked the ELAIS calibration using some dedicated
measurements of standard stars and ELAIS sources. We claim that the MPIA ELAIS fluxes are calibrated
with an accuray better than 20\%. We listed 274 reliable detections and 642 probable detections.
A 2D and 3D spatial correlation analysis is on-going.      

\section*{ACKNOWLEDGMENTS}
C. S. would like to thank U. Herbstmeier for his contribution to this study,
the ISOPHOT group in Heidelberg, C. Gabriel for valuable discussions and
 B. Schulz for providing calibration data.


\begin{thebibliography}{}

\bibitem[\protect\astroncite{Ciliegi et~al.}{1998}]{pc98}
Ciliegi P. and the ELAIS consortium 1998,
MNRAS, in press

\bibitem[\protect\astroncite{Gabriel et~al.}{1996}]{cg96}
Gabriel C. et al., 1996, 
PIA Users Manual, 
available from ESA/ISO Ground Observatory VILSPA or MPIA Heidelberg

\bibitem[\protect\astroncite{H\'eraudeau et~al.}{1999}]{ph99}
H\'eraudeau P., Surace C., Lemke D., Oliver S., Rowan-Robinson M.,
in "The Universe as seen by ISO"

\bibitem[\protect\astroncite{Oliver et~al.}{1997}]{so97}
     Oliver S. and the ELAIS consortium , 1997
     in "New Horizons from Multi-Wavelength Sky
     Surveys", Ed: Brian McLean Pubs: Kluwer 
     IAUS 179, 112

\bibitem[\protect\astroncite{Oliver et~al.}{1998}]{so98}
Oliver, S., and the ELAIS consortium 1998,
MNRAS, in press

\bibitem[\protect\astroncite{Rowan-Robinson et~al.}{1999}]{mrr99}
 Rowan-Robinson M. and the ELAIS consortium, 
in "The Universe as seen by ISO"

\bibitem[\protect\astroncite{Surace et~al.}{1997}]{cri97}
Surace C., Abraham P., Herbstmeier U., ' New flatfielding correction method for ISOPHOT C100/C200 detector', June
     1997, ISOPhot Interactive Analysis reports

\bibitem[\protect\astroncite{Schultz}{private communication}]{bs98}
Schultz, B., private communication 
\end{thebibliography}
\end{document}